# Smith-Purcell radiation from a charge moving above a finite-length grating with rectangular profiles

Weiwei Li[1*], Weihao Liu[2], Zhigang He[1], Qika Jia[1], and Lin Wang[1]

[1]National Synchrotron Radiation Laboratory, University of Science and Technology of China, Hefei, Anhui, 230029, China

[2]College of Electronic and Information Engineering, Nanjing University of Aeronautics and Astronautics, Nanjing, Jiangsu, 211106, People's Republic of China

(Dated: 12 April 2020)

**Abstract**

Smith-Purcell radiation is generated by a charged particle beam passing close to the surface of a diffraction grating which has a strong dependency of the emitted radiation intensity on the form of the grating profile. For relativistic electron beam, it is important to take into account the number of grating periods in practical SPR setups. In this paper, the theoretical investigations of Smith-Purcell radiation from a three-dimensional bunch of relativistic electrons that moves at constant speed parallel to an electrically perfectly conducting grating with finite rectangular grooves are carried out by using the modal matching method. This model may offer a new efficient tool for terahertz production by SPR interaction and for nondestructive bunch-length measurements by SPR.

**Introduction**

Smith-Purcell radiation[1] is originated when a charged particle travels parallel to a plane with diffraction grating. Recent renewed interest in this problem is caused by different applications. Among these applications are length determination for short electron bunches[2-3], creation of monochromatic light source in the various spectral regions[4-5].

Various theoretical models were proposed for describing the SPR. However, the general theory of SPR has not been created yet and at present there exist a number of different approaches, which sometimes are in agreement and sometimes contradict each other[2,5]. The rigorous solution by van den Berg of the SPR from an infinitely long grating is obtained by solving an integral equation having a periodic Green's function[6]. For the non-relativistic electron energies, the approach developed by van den Berg ensures a reasonable agreement with experiments. In principle, an arbitrary tooth profile can be analyzed with this approach. In practice, extensive numerical computation is generally required to approach the asymptotic value for the radiation intensity, however for particular case of rectangular grating, the modal matching method is recommended since it agrees with integral method, but takes shorter computational time[7].

For relativistic electron beam, it is important to take into account the number of grating periods in practical SPR setups. A model based on the finite-difference time-domain (FDTD) method, in which a total-field scattered-field technique combined with a near- to far-field projection, is described in[8]. A frequency-domain model based on the electric-field integral equation (EFIE)[9] extends van den Berg's model for practical gratings of finite length and agrees with the FDTD model. Both models are adaptable to arbitrary grating geometries but requires extensive numerical computation. Accurate far-field measurements of SPR with a 15 MeV beam have demonstrated good agreement between measured power and the predictions of EFIE theory[10].

---

* email address: liwe@ustc.edu.cn

In this paper, we will develop the modal matching method for the calculations of finite-grating with rectangular grooves, which is one of the most popular grating types. Although the EFIE theory is also still applicable for this case, the modal matching method will take much less computation. In the previous theoretical work [7,11,12] which applied the modal matching method on SPR from the rectangular gratings, the periodic grating conditions are used, but operating the SPR under the periodic regime driving by the relativistic electrons would require a large number of grating grooves which is difficult to fulfill in the practical experiments.

## Formula Derivation

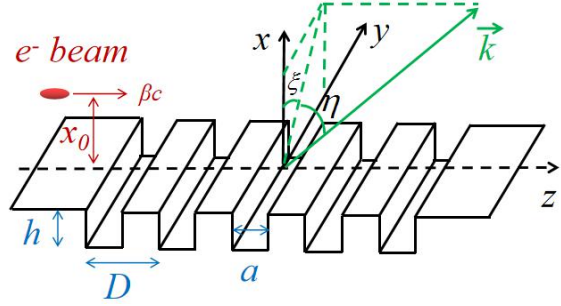

Fig. 1 The SPR scheme. An electron bunch is moving above a grating with rectangular grooves

In the formula derivations, we primarily consider the case that the electron beam excites an electrically perfectly conducting grating with finite rectangular grooves, and the schematic diagram of which is illustrated in Fig. 1.

The incident filed can be given by the Fourier transform of the free space field produced by an electron beam with charge of $Q$ moving along the $z$ direction at a relativistic velocity $\beta c$ [6]

$$H_y^{in}(x,z;k_y,\omega) = -\frac{Q}{2} F_{coh} \text{sign}(x-x_0) \exp(jk_{z0}(z-z_0)) \exp(jk_{x0}|x-x_0|)$$
$$E_y^{in}(x,z;k_y,\omega) = \frac{Q}{2} F_{coh} \sqrt{\frac{\mu_0}{\varepsilon_0}} \frac{k_y}{k_0} \frac{k_{z0}}{k_{x0}} \exp(jk_{z0}(z-z_0)) \exp(jk_{x0}|x-x_0|)$$

(1)

where $x_0$ and $z_0$ are the initial charge coordinates, $k_{z0} = k/\beta$ and $k_{x0} = \sqrt{k^2 - k_{z0}^2 - k_y^2}$ and the $F_{coh}$ is the coherence bunch factor. Assuming uncorrelated electron distributions, $\rho(x,y,t) = X(x)Y(y)T(t)$, the coherent integral is given by [2]

$$F_{coh} = \int_{-\infty}^{+\infty} X(x)\exp(jk_{x0}(x-x_0))dx \int_{-\infty}^{+\infty} Y(y)\exp(jk_y y)dy \int_{-\infty}^{+\infty} T(t)\exp(j\omega t)dt \qquad (2)$$

The reflected waves in the region above the grating can be expanded in continuous series,

$$E_y^r(x,z;k_y,\omega) = \frac{1}{2\pi} \int_{-\infty}^{+\infty} dk_z E_{y\omega}^r(k_z,k_y,\omega) \exp(jk_z z + jk_x x)$$
$$H_y^r(x,z;k_y,\omega) = \frac{1}{2\pi} \int_{-\infty}^{+\infty} dk_z H_{y\omega}^r(k_z,k_y,\omega) \exp(jk_z z + jk_x x)$$

(3)

The radiation directions can be defined in terms of the angular coordinate in relation to the wave-numbers

$$k_z = k_0 \sin\eta; k_y = k_0 \cos\eta \sin\zeta; k_x = k_0 \cos\eta \cos\zeta \qquad (4)$$

The field in the grooves can be expanded into a series of open mouth waveguide modes [11]

$$\begin{cases} H_{\omega,y}^{II} = \sum_{m=0}^{\infty} B_{pm} \cos\left(\frac{m\pi}{a_p}(z-z_p)+\frac{m\pi}{2}\right)\left[\exp(-j\kappa_{pm}x)+\Gamma_{pm}\exp(j\kappa_{pm}x)\right] \\ E_{\omega,y}^{II} = \sum_{m=1}^{\infty} C_{pm} \sin\left(\frac{m\pi}{a_p}(z-z_p)+\frac{m\pi}{2}\right)\left[\exp(-j\kappa_{pm}x)-\Gamma_{pm}\exp(j\kappa_{pm}x)\right] \\ \kappa_{pm} = \sqrt{k_0^2 - k_y^2 - \left(\frac{m\pi}{a_p}\right)^2}\,;\,\Gamma_{pm} = \exp(j2\kappa_{pm}h_p);\ z_p - \frac{a_p}{2} \leq z \leq z_p + \frac{a_p}{2} \end{cases} \quad (5)$$

where $z_p$, $a_p$ and $h_p$ are the center longitudinal position, width and depth of $p$-th groove respectively. The number of the grooves is $N_g$. It should be noted that the groove sizes can be set to be different from each other.

Using boundary conditions on the grating and continuity conditions between the solutions of the upper half space and those in the grooves, we get a set of algebraic equations for the unknowns coefficients. The system for the unknown waveguide amplitude are

$$B_{pm}\frac{1+\delta_{m0}}{2}(\Gamma_{pm}+1) - \frac{1}{2\pi}\int_{-\infty}^{+\infty} dk_z \frac{\sum_{q=1}^{N_g}\sum_{n=0}^{\infty} a_q \kappa_{qn} B_{qn}(\Gamma_{qn}-1)}{k_x}\Phi(q,n,k_z)\Phi^*(p,m,k_z) = H_0\Phi^*(p,m,k_{z0}) \quad (6)$$

and

$$\frac{1}{2}C_{pm}\kappa_{pm}(1+\Gamma_{pm}) + \int_{-\infty}^{+\infty} dk_z \frac{k_x}{2\pi} \sum_{q=1}^{N_g}\sum_{n=1}^{\infty} C_{qn} a_q \Psi(q,n,k_z)[1-\Gamma_{qn}]\Psi^*(p,m,k_z) = E_0 k_{x0}\Psi^*(p,m,k_{z0}) \quad (7)$$

with

$$\begin{cases} H_0 = QF_{coh}\exp(-jk_{z0}z_0+jk_{x0}x_0) \\ E_0 = QF_{coh}\sqrt{\frac{\mu_0}{\varepsilon_0}}\frac{k_y}{k_0}\frac{k_{z0}}{k_{x0}}\exp(-jk_{z0}z_0+jk_{x0}x_0) \\ \Phi(p,m,k_z) = \exp(-jk_z z_p)\frac{1}{a_p}\int_{-a_p/2}^{a_p/2}\exp(-jk_z z)\cos\left(\frac{m\pi z}{a_p}+\frac{m\pi}{2}\right)dz \\ \Psi(p,m,k_z) = \exp(-jk_z z_p)\frac{1}{a_p}\int_{-a_p/2}^{a_p/2}\exp(-jk_z z)\sin\left(\frac{m\pi}{a_p}z+\frac{m\pi}{2}\right)dz \end{cases} \quad (8)$$

The coefficients of the field in the upper half space can be then obtained by

$$\begin{cases} -\pi k_{x0}H_0\delta(k_{z0}-k_z) + k_x H_{y\omega}^r = \sum_{p=1}^{N_g}\sum_{m=0}^{\infty}\kappa_{pm}B_{pm}a_p\Phi(p,m,k_z)\{-1+\Gamma_{pm}\} \\ \pi E_0\delta(k_{z0}-k_z) + E_{y\omega}^r = \sum_{p=1}^{N_g}\sum_{m=1}^{\infty} C_{pm}a_p\Psi(p,m,k_z)[1-\Gamma_{pm}] \end{cases} \quad (9)$$

The radiated energy related to the amplitudes of the reflected waves can be expressed as

$$\frac{dW}{d\omega d\Omega} = \frac{1}{4\pi^3}\frac{k_x^2 c(\mu_0 H_{y\omega}^2 + \varepsilon_0 E_{y\omega}^2)}{1 - k_y^2/k_0^2} \quad (10)$$

## Summary

In this paper, we have carried out the modal matching method for the calculations of finite-grating with rectangular grooves. This method can be an alternative and equivalent method of the EFIE theory for the specific situation. More characteristics about the SPR from finite-grating with rectangular grooves will be studied in the future work.

**Acknowledgements**

This work is supported by National Foundation of Natural Sciences of China (11705198)